\documentclass[a4paper,12pt]{article}
\renewcommand{\baselinestretch}{1.2}
\begin{document}
%
%\setlength{\textwidth}{18cm}
%\large
%
\title{Probing the mathematical nature of the photon field}
\author{Walter Smilga \\ Isardamm 135 d, D-82538 Geretsried, Germany \\
e-mail: wsmilga@compuserve.com}
\maketitle 
\renewcommand{\baselinestretch}{1.2}

\begin{abstract}
The mathematical content of the interaction term of quantum 
electrodynamics is examined under the following assumption: 
It is presumed that the apparent degrees-of-freedom of 
the photon field reflect the kinematical degrees-of-freedom of
the two-particle state space of massive fermions, rather than
independent degrees-of-freedom of the photon field.
This assumption is verified by reproducing the numerical value 
of the fine-structure constant.
\end{abstract}

\renewcommand{\baselinestretch}{1.0}

\section{Introduction}

The development of quantum electrodynamics (QED) ranks among the 
grea\-test success stories of theoretical physics. 
Calculations within QED agree with experimental data to any required degree 
of precision.
QED therefore has served generations of physicists as a guideline in extending
its methods to other interactions.  
The suggestive power of Feynman diagrams brought forward such fascinating 
concepts as virtual particles, virtual pair production, vacuum fluctuation 
and off-shell behaviour, 
which have dominated our understanding of elementary processes until the 
present day. 

As a matter-of-fact the computation method, visualized by Feynman diagrams,
is first and foremost a mathematical algorithm. 
The success of QED is owed to this algorithm, but not to its physical 
interpretation in terms of virtual particles and vacuum fluctuations.

The algorithm of QED is based on a multi-particle representation in Fock 
space.
Creation and destruction operators are used as mathematical tools to 
construct particle states from a ``vacuum" state. 
This ``vacuum" state is a mathematical construct that has nothing to do 
with a physical vacuum. 
Creation and destruction operators, despite their names, have nothing in 
common with physical ``creation" or ``destruction" of particles. 
Internal lines in Feynman diagrams are the result of contracting a pair of 
creation and destruction operators. 
Contraction results in C-numbers, which again have nothing in common 
with real or virtual particles. 
Therefore, from a mathematical point of view there is no reason to associate 
internal lines with particles, either physically or virtually.

Of course, we are free (and actually it may be helpful to a certain extent), 
to underlay an algorithm with appropriate images, as long as we keep in mind 
that these are images only, which may or may not have a correspondence to 
physical reality.
Unfortunately, an understanding based on images may obstruct our view of 
deeper interrelations and prevent us from asking the right continuative 
questions. 

In view of the highly dynamic image of a fluctuating vacuum and photons, 
permanently transforming to particle-antiparticle pairs and back to photons, 
the following question may appear absurd:
Could it be that the quantized electromagnetic field does not possess 
degrees-of-freedom on its own, but rather reflects degrees-of-freedom of 
the field generating charged particles?

As a matter of fact, Feynman in one of his seminal papers on quantum 
electrodynamics \cite{rpf} completely eliminated photons and developed the 
algorithm, which he visualized by Feynman diagrams, from a direct 
action at a distance between charges. 
Therefore, the question is not as absurd as it may appear from the
electromagnetic field point of view.

This paper will give an answer to the question, whether the quantized
electromagnetic field holds degrees-of-freedom on its own, or merely reflects 
kinematical degrees-of-freedom of the field generating particles.
As a by-product a numerical value of the fine-structure constant $\alpha$
will be deduced from geometric properties of the two-particle state space 
of massive fermions.

\section{Counting degrees-of-freedom}

The interaction term of QED in momentum space essentially consists 
of terms with the structure (cf. e.g. \cite{sss})
\begin{equation}
\dots\; {\bar{b}({\mathbf{p + k}}) \, \gamma_\mu \, b({\mathbf{p}})} \, 
a^\mu({\mathbf{k}}) \dots  \; .                           \label{1-1}
\end{equation}
The interpretation of such a term is: a photon with momentum ${\mathbf{k}}$, 
represented by the destruction operator $a^\mu({\mathbf{k}})$, is absorbed 
and the momentum of an electron, represented by destruction and creation 
operators ${b({\mathbf{p}})}$ and ${\bar{b}({\mathbf{p + k}})}$, is changed 
from ${\mathbf{p}}$ to ${\mathbf{p + k}}$.

Electron-electron scattering is described by two such terms,
with an integration over the photon momenta ${\mathbf{k}}$ and 
${\mathbf{k'}}$,
\begin{equation}
\int d{\mathbf{k}}\, d{\mathbf{k'}}
\dots\; {\bar{b}({\mathbf{p + k}}) \, \gamma_\mu \, b({\mathbf{p}})} \, 
a^\mu ({\mathbf{k}})  \dots \;\;                            
\dots\; {\bar{b}({\mathbf{p' - k'}}) \, \gamma_\nu \, b({\mathbf{p'}})} \, 
a^\nu({\mathbf{k'}})^\dagger \dots \; .                       \label{1-2}
\end{equation}

This expression represents a Fock space operator, acting on the 
photon state space (and, of course, also on the electron state space).
As such it must offer \begin{em}exactly one\end{em} destruction and/or 
creation operator for each ``independent" photon state. 
Photon states are characterized by their polarization and by their momentum.
In the perturbation calculation of QED the photon creation and destruction 
operators are, therefore, found in integrals over the full photon 
momentum space.  
 
The following will be based on the working hypothesis, that a photon is
not a ``physical" entity, with its own independent degrees-of-freedom, but 
rather an auxiliary mathematical construct.
This construct is used to describe the transition of a two-particle state 
with a given total momentum to another two-particle state with the same 
total momentum.
The transition can still be described by the ``exchange of a photon", but 
it is presumed that the role of ``independent" states is taken over by 
two-particle states.

Accordingly, in (\ref{1-3}) the integration over ${\mathbf{k}}$ and 
${\mathbf{k'}}$ is replaced by an integral over those 
two-particle states that contribute to the scattering amplitude.
The integration is over the particle momenta ${\mathbf{p'}}$ and 
${\mathbf{q'}}$ located in a domain ${\mathcal{P}}$ that corresponds to 
the contributing ``outgoing" two-particle states. 
\begin{equation}
\int_{\mathcal{P}} d{\mathcal{P}}({\mathbf{p'}},{\mathbf{q'}})
\dots\; {\bar{b}({\mathbf{p'}}) \, \gamma_\mu \, b({\mathbf{p}})} \, 
a^\mu ({\mathbf{p' - p}})  \dots \;\;                            
\dots\; {\bar{b}({\mathbf{q'}}) \, \gamma_\nu \, b({\mathbf{q}})} \, 
a^\nu({\mathbf{q - q'}})^\dagger \dots  \; .                 \label{1-3}
\end{equation}

If \begin{em}all\end{em} two-particle states that can be built from two 
single-particle states would contribute, then ${\mathcal{P}}$ would be
identical to $R^3 \times R^3$ and the integral would be replaceable by
\begin{equation}
\int_{R^3 \times R^3} d{\mathbf{p'}} d{\mathbf{q'}} \dots \; . \label{1-3a}
\end{equation}
This would mean only a reordering of the integrations in (\ref{1-2}).
However, when the 4-momentum of the incoming two-particle state is given,
an irreducible two-particle representation of the Poincar\'e group is fixed.
As long as the 4-momentum of the two-particle state is conserved, the outgoing 
two-particle state must belong to the same representation.
This suggests to include into ${\mathcal{P}}$ only such two-particle 
states that belong to the same irreducible representation of the Poincar\'e
group. 

Obviously, the parameter space ${\mathcal{P}}$ is a subspace of 
$R^3 \times R^3$.
Therefore, there exists a representation of the integral in the form
\begin{equation}
\int_{R^3 \times R^3} W({\mathbf{p'}},{\mathbf{q'}}) \,
d{\mathbf{p'}} d{\mathbf{q'}} \dots \; ,                \label{1-3b}
\end{equation}
where 
\begin{equation}
W({\mathbf{p'}},{\mathbf{q'}}) = 
\frac{\partial{\mathcal{P}}({\mathbf{p'}},{\mathbf{q'}})}
{\partial{\mathbf{p'}} \partial{\mathbf{q'}}}           \label{1-3c}
\end{equation}
is the Jacobian obtained from the parameterisation of the (non-Euclidean) 
parameter space ${\mathcal{P}}$ by Euclidean parameters.
$W$ is a measure for the reduction of degrees-of-freedom within the
outgoing states, caused by the restriction to an irreducible two-particle
representation.

A closer look at this Jacobian will help to decide, whether the apparent 
degrees-of-freedom of the photon field belong to the photon field itself, 
or to the outgoing two-particle states.

The determination of the Jacobian will be based on geometric properties of
the momentum parameter space of the contributing two-particle states. 
It will result in a step-wise combination of some characteristic symmetric 
structures, which will finally determine $W({\mathbf{p'}},{\mathbf{q'}})$.

In fact, this task was (unwittingly) tackled 40 years ago by 
A. Wyler \cite{aw}, who discovered that the fine-structure constant $\alpha$ 
can be expressed by volumes of certain symmetric spaces. 
Unfortunately, Wyler was not able to put his observation into a convincing 
physical context. 
Even worse, his mathematical reasoning did not withstand a closer 
inspection.
Therefore, his work was criticized as fruitless numerology \cite{br}.
 
To obtain information about the domain ${\mathcal{P}}$, the parameter space 
of two-particle states will be examined.
The state of a single particle is determined by three independent
parameters $p_1,p_2,p_3$ and the spin.
The parameters $p$ satisfy the relation
\begin{equation}
p_0^2 -p_1^2 - p_2^2 - p_3^2 = m^2 , \label{10-1}
\end{equation} 
where $m$ is the particle mass.
The independent parameters $p_1,p_2,p_3$ span a 3-dimensional 
parameter space. 
Given a state of the state space, then the full state space and the
corresponding parameter space are generated by application of the 
symmetry group SO(3,1).

Two-particle states are consequently described by a 6-dimensional 
parameter space.
The parameter space generating group SO(3,1)$\times$SO(3,1) 
leaves the relation 
\begin{equation}
p_0^2 + p_0^{'2} - p_1^2 - p_2^2 - p_3^2 
- p_1^{'2} - p_2^{'2} - p_3^{'2} = 2\,m^2     \label{10-2}
\end{equation} 
invariant.

When the momenta of both particles are given, an effective mass $M$ of 
the two-particle state can be calculated 
\begin{equation}
(p_0+p_0')^2 - (p_1+p_1')^2 - (p_2+p_2')^2 - (p_3+p_3')^2 = M^2 
\;.                                            \label{10-2a}
\end{equation}
The value of $M$ characterizes an irreducible representation of
the Poincar\'e group P(3,1). 
When $M$ is hold fixed, (\ref{10-2}) and (\ref{10-2a}) result in
\begin{equation}
p_0\,p_0' - p_1\,p_1' - p_2\,p_2' - p_3\,p_3' = const \;. \label{10-2b}
\end{equation}
This condition reduces the number of independent parameters to five.
In the rest frame of the primed particle (\ref{10-2b}) takes on the
simple form
\begin{equation}
p_0 = const  \label{10-2c}
\end{equation}
and with (\ref{10-1}) we have
\begin{equation}
p_1^2 + p_2^2 + p_3^2 = const \;. \label{10-2d}
\end{equation} 
If we choose $p_1$ and $p_2$ as independent parameters, then these parameters
fill the disk inside and including the circle 
\begin{equation}
p_1^2 + p_2^2 = const \;. \label{10-2e}
\end{equation}
Therefore, the parameter space of an irreducible two-particle state space can 
be gene\-rated by the symmetry group SO(3,1)$\times$SO(2,1).
This symmetry group is a subgroup of SO(5,2).

Associated with SO(5,2) is the quotient group SO(5,2)/(SO(5)$\times$SO(2)).
A realization of this quotient group is the irreducible homogeneous bounded 
symmetric domain (cf. \cite{lkh}) of 5-dimensional complex vectors $z$ 
\begin{equation}
D^5 = \{z \in C^5; 1 + |zz'|^2-2\bar{z}z' > 0, |zz'| < 1 \}\;. \label{1-4}
\end{equation}
The boundary of $D^5$ is given by
\begin{equation}
Q^5 = \{\xi = x\,e^{i\theta}; x \in R^5, xx'=1 \},\;
\mbox{with}\;0<\theta<\pi\;.                                   \label{1-6}
\end{equation}
SO(5)$\times$SO(2) acts transitively on $Q^5$.
Therefore $Q^5$ serves as a natural parameterisation of any irreducible state 
space that is invariant under the symmetry group SO(5)$\times$SO(2).
Below, $D^5$ will be used to extend this parameterisation to a state space 
that is invariant under (a subgroup of) SO(5,2).
Irreducible homogeneous bounded symmetric domains were extensively stu\-died
by L. K. Hua\cite{lkh}, who also calculated some relevant volumes. 

Let us now apply this parameterisation to the integral (\ref{1-3}) and start
with the integration over $Q^5$ 
\begin{equation}
\int_{Q^5} \frac{d\xi}{V(Q^5)} \dots \;.      \label{1-9}
\end{equation}
This means, in a first step we integrate over parameters that are located 
on $Q^5$.
The Jacobian $V(Q^5)^{-1}$ is the inverse surface volume of $Q^5$.
The 4-dimensional infinitesimal volume element in (\ref{1-9}) already has 
the form of a volume element in Cartesian coordinates.
In a second step we will adjust this volume element to the parameter 
space associated with the subgroup SO(3)$\times$SO(2) of SO(5).
In a third step we will extend it to a 5-dimensional Cartesian volume element,
which will cover also the boost-operations of SO(3,1)$\times$SO(2,1).

The integral (\ref{1-9}) ranges over the full SO(5)-symmetric $Q^5$.
As said before, the two-particle state space is not fully SO(5) symmetric: 
A rotation from the momentum subspace of the first particle to 
the second is not a valid symmetry operation and hence cannot
contribute to the parameter space.
Excluding this axis of rotation from the symmetry operations
of $Q^5$ means a reduction of the integration volume by a factor 
determined from the quotient group SO(5)/SO(4).
This quotient group is isomorphic to the unit sphere $S^4$ in 
five-dimensional Euclidean space $R^5$.
Therefore, the volume of the parameter space and consequently the
integration volume has to be corrected by a factor of 
$1/V(S^4)$ applied to the volume element $d\xi$.
This adds the factor $1/V(S^4)$ to the Jacobian in (\ref{1-9}).

The integration over the phase $\theta$ (cf. (\ref{1-6})) delivers a factor
of $2\pi$, because the integrand in (\ref{1-3}) does not explicitly depend 
on $\theta$.
(Phase $\theta$ corresponds to the orientation of the two-particle 
energy-momentum vector in the $p'_0$-$q'_0$-plane,
but $p'_0$ and $q'_0$ can be eliminated by using relation (\ref{10-1}) ).
Then, from now on, $\xi$ can be considered as real. 

To extend the infinitesimal volume element $d\xi$ to an 
\begin{em}Euclidean\end{em} infinite\-simal volume element in 
\begin{em}five\end{em} dimensions, we need another integration in a 
radial direction perpendicular to a surface element of $Q^5$.
Let $\xi_5$ be the parameter of this direction.
The extended volume element shall be \begin{em}isotropic\end{em} with 
respect to all five directions.
This requires some additional thoughts.

Consider the formula that relates the volume of a polydisk $D^5_R$ with 
radius $R$ to the volume of the unit polydisk (\ref{1-4})
\begin{equation}
V(D^5_R) = R^5 \, V(D^5) \; .
\end{equation}
This volume can be written as an integral over five independent 
radii $r$, $r'$, $r''$, $r'''$, $r''''$ 
\begin{equation}
5 \int^R_0 dr 
\int^r_0 V(D^5)^{\frac{1}{4}} dr'
\int^r_0 V(D^5)^{\frac{1}{4}} dr''
\int^r_0 V(D^5)^{\frac{1}{4}} dr'''
\int^r_0 V(D^5)^{\frac{1}{4}} dr''''  \; . \label{1-10}
\end{equation}
Another representation of the volume of $D^5_R$ is given by the formula
\begin{equation}
V(D^5_R) = 5 \int^R_0 dr \int_{Q^5} r^4 \, dQ^5 \; .
\end{equation}
A comparison with (\ref{1-10}) shows that the contributions of the four 
integrals with boundaries $0$ and $r$ are directly related to the integral 
over the four-dimensional volume element of $Q^5$.
Therefore, each of the four directions of $dQ^5$ contributes to the volume of 
$D^5_R$ with a factor $V(D^5)^{\frac{1}{4}}$.
When we extend the integration in (\ref{1-9}) into the direction 
perpendicular to a given surface element of $Q^5$, and demand isotropy with 
respect to all five directions, then we must scale the infinitesimal volume 
element for the fifth direction with the same factor.
The infini\-tesimal volume element for the integration in the direction 
of $\xi_5$ is therefore given by
\begin{equation}
V(D^5)^\frac{1}{4}\, d\xi_5 \; .  \label{1-11}
\end{equation}  

The completed volume element can be used to integrate over the full
Eucli\-dean $R^5$.

Remembering that $2 \times 2$ spin components contribute to a two-particle 
state, we add to the Jacobian a factor of 4.

Collecting all factors results in an effective volume factor
\begin{equation}
W = 8 \pi \,V(D^5)^{\frac{1}{4}} \, / \, (V(S^4) \, V(Q^5)) \;.  \label{1-12}
\end{equation}
This is Wyler's formula.
In collecting the contributions to $W$ we have not found any dependency
on the momentum parameters.
Therefore, $W$ is a constant.

With respect to the outgoing two-particle states in (\ref{1-3}), this formula 
delivers the number of independent degrees-of-freedom, associated with the 
volume element $d{\mathcal{P}}({\mathbf{p'}},{\mathbf{q'}})$, relative to
those of a 5-dimensional Euclidean volume element $d^5p$.
Wyler's formula acts as a Jacobian $W$ that allows us to rewrite 
the integral over ${\mathcal{P}}$ as an integral over the Euclidean 
parameter space $R^5$:
\begin{equation}
\int_{\mathcal{P}} d{\mathcal{P}}({\mathbf{p'}},{\mathbf{q'}}) \dots 
\;\;\;\rightarrow\;\;\;  
\int_{R^5} W \; d^5p \dots \;.  \label{1-13}
\end{equation}

When we identify differences of $p$ with photon momenta, we are almost back 
to the form of integral (\ref{1-2}).
The integration in (\ref{1-2}) is over $R^6$, whereas the last integral in 
(\ref{1-13}) is over $R^5$.
However, we can (and have to) extend the integration in (\ref{1-13}) to an 
integral over $R^6$
\begin{equation}
\int_{R^5} W \; d^5p \dots \;
\;\;\;\rightarrow\;\;\;  
\int_{R^6} W \; d{\mathbf{k}}\, d{\mathbf{k'}} \dots \;,  \label{1-14}
\end{equation}
since the Feynman-Dyson perturbation expansion provides $\delta$-functions
co\-ve\-ring all 6 parameters. 
The integration over the $\delta$-function, belonging to the additional 
parameter, delivers a factor of 1, since the other parts of the integrand do 
not depend on this parameter.
We obtain
\begin{equation}
\int W \, d{\mathbf{k}}\, d{\mathbf{k'}}
\dots\; {\bar{b}({\mathbf{p + k}}) \, \gamma_\mu \, b({\mathbf{p}})} \, 
a^\mu ({\mathbf{k}})  \dots \;\;                            
\dots\; {\bar{b}({\mathbf{p' - k'}}) \, \gamma_\nu \, b({\mathbf{p'}})} \, 
a^\nu({\mathbf{k'}})^\dagger \dots  \; .                    \label{1-15}
\end{equation}
The only difference to (\ref{1-2}) is the factor $W$, which results
from treating the states of an irreducible two-particle representation, 
rather than photon states, as independent degrees-of-freedom.
The factor $W$ turns up in the position of the squared 
coupling constant between fermion and photon field.
In the standard treatment of QED, in this position the empirical
value of the fine-structure constant is inserted ``by hand".
 
The volumes $V(D^5)$ and $V(Q^5)$ have been calculated by 
L. K. Hua \cite{lkh}. 
$V(S^4)$ is the volume of the unit sphere in 4 dimensions. With
\begin{equation}
V(Q^5) = \frac{8 \pi^3}{3},        			\label{1-16}
\end{equation}
\begin{equation}
V(D^5) = \frac{\pi^5}{2^4\, 5!},   			\label{1-17}
\end{equation}
\begin{equation}
V(S^4) = \frac{8 \pi^2}{3}         			\label{1-18}
\end{equation}
we obtain 
\begin{equation}
\frac{9}{8 \pi^4} \left(\frac{\pi^5}{2^4 \, 5!}\right)^{1/4}   	
 = \; \frac{9}{16 \pi^3} \left(\frac{\pi}{120}\right)^{1/4} 
= \; 1/137.03608245.   			          	\label{1-19}
\end{equation}
The best experimental (low energy) value for $\alpha$, determined from 
the electron magnetic moment, currently is 1/137.035 999 084(51) \cite{hfg}.

\section{Conclusion}

A geometric measure for the number of independent two-particle states in a 
matrix element of electron-electron scattering has been obtained.
It enters the matrix element at the position of the coupling 
constant and shows an excellent agreement with the experimental (low-energy) 
value of $\alpha$. 
It leaves a fingerprint that is uniquely related to the degrees-of-freedom of 
an irreducible two-particle representation.
The coincidence with $\alpha$ can, therefore, be considered as an experimental 
indication that the low-energy behaviour of the electromagnetic interaction is 
determined by contributions from states of a single irreducible 
two-particle representation.
This may not come as a surprise, as far as the lowest order of the perturbation
series is concerned.
But this statement applies to all orders, since all terms contain the same 
value of $\alpha$.
This result shows evidence that the photon field does not 
possess degrees-of-freedom of its own. 
It rather relays the kinema\-tical degrees-of-freedom of the charged particles 
that are responsible for the photon field.  

Reassigning the degrees-of-freedom from photon states to two-particle 
states has a serious impact on our understanding of the photon field.
Never\-the\-less, it has no influence on the perturbation algorithm.
Therefore, all calculations of QED remain unaffected. 
This implies that the reinterpretation of the photon field is in 
agreement with experimental data.

Although the numerical agreement of $W$ and $\alpha$ is more than sufficient
to answer the question of degrees-of-freedom, 
it cannot be denied that the value of $W$ differs from the experimental value 
of $\alpha$ by an amount that is significantly larger than the present 
uncertainty of $\alpha$.
This indicates that the number of independent states is slightly 
larger than calculated by a factor of roughly 1.0000005.
At higher energies considerably larger corrections to the low-energy coupling 
constant apply (running fine-structure constant, cf. e.g. \cite{fj}).

There is a simple explanation for this behaviour.
The derivation of Wyler's formula has been based on the presumption that the 
representation of two-particle states is conserved during the scattering 
process.
When this presumption ceases to apply, e.g. due to the generation of 
outgoing high-energy photons, then an integration over different 
representations will be required.
In the extreme case this may result in an integral of the form (\ref{1-3a}).
The coupling constant, determined from the Jacobian of this integral, has 
the value 1; it can be understood as a ``high-energy" limit of the coupling 
constant.
Obviously, Wyler's formula relates to the opposite, ``low-energy"
limit, where the energy of outgoing photons approaches zero.

\renewcommand{\baselinestretch}{1.1}


\begin{thebibliography}{99}

\bibitem{rpf} R. P. Feynman, 
Phys. Rev. \bfseries76\mdseries, 769-789 (1949).

\bibitem{sss} G. Scharf, \begin{em}Finite Quantum Electrodynamics\end{em}, 
(Springer, Berlin Heidelberg New York, 1989).

\bibitem{lkh} L. K. Hua, \begin{em}Harmonic Analysis of Functions of
Several Complex Variables in the Classical Domains\end{em},
(American Mathematical Society, Provi\-dence, 1963).

\bibitem{aw} A. Wyler,
C. R. Acad. Sc. Paris \bfseries271A\mdseries, 186-188 (1971).

\bibitem{br} B. Robertson, 
Phys. Rev. Lett. \bfseries27\mdseries, 1545 (1971).

\bibitem{hfg} D. Hanneke, S. Fogwell, G. Gabrielse,
Phys. Rev. Lett. \bfseries100\mdseries, 120801 (2008).

\bibitem{fj} F. Jegerlehner,
Nucl. Phys. B - Proceedings Supplements, \bfseries181-182\mdseries, 
135-140 (2008); arXiv:0807.4206[hep-ph]

\end{thebibliography}
\end{document}